\documentstyle[aps,multicol,prl,psfig,epsf]{revtex}

\begin{document}
\title
{
Canonical phase space approach to the noisy Burgers equation
}
\author{Hans C. Fogedby}
\address{
\thanks{Permanent address}
Institute of Physics and Astronomy,
University of Aarhus, DK-8000, Aarhus C, Denmark
\\
and
NORDITA, Blegdamsvej 17, DK-2100, Copenhagen {\O}, Denmark
}
\maketitle
\begin{abstract}
Presenting a general phase approach to stochastic processes we analyze 
in particular 
the  Fokker-Planck equation for the noisy Burgers equation 
and discuss the time dependent and stationary probability distributions.
In one dimension we derive the long-time skew distribution 
approaching the symmetric stationary Gaussian distribution.
In the short time regime we discuss heuristically the nonlinear
soliton contributions and derive an expression for the distribution
in accordance with the directed polymer-replica  model
and asymmetric exclusion model results.
\end{abstract}
\draft
\begin{multicols}{2}
\narrowtext
The strong coupling aspects of systems driven stochastically
far from equilibrium present a formidable challenge in modern
statistical physics and soft condensed matter.
The phenomena in question are ubiquitous and include turbulence in
fluids, interface and growth problems, chemical reactions, self organized
critical systems, and even economical and sociological models.

In recent years much of the focus of modern statistical physics
and soft condensed matter has shifted towards such systems. Drawing
on the case of static and dynamic critical phenomena in and close
to equilibrium, where scaling, critical exponents, and universality
have served to organize our understanding and to provide calculational
tools, a similar approach  has been advanced towards nonequilibrium
phenomena with the purpose of elucidating scaling properties and
more generally the morphology or pattern formation in a driven state.

Whereas perturbative field theory together with the dynamic renormalization
group have proven successful in the context of dynamic critical phenomena,
the extension to nonequilibrium phenomena is often plagued with both technical
and conceptual problems. This is related to the occurrence of 
strong coupling features encountered for example in the notable case of
hydrodynamical turbulence, and there is a need for the development of
appropriate nonperturbative theoretical tools in order to access the
strong coupling regime.

In this context the noisy Burgers equation for the slope
$u_n=\nabla_nh$ ($n=1,..d$) of a growing interface \cite{1},
\begin{equation}
\frac{\partial u_n}{\partial t}=\nu{\nabla}^2u_n
+\lambda u_p\nabla_pu_n + \nabla_n\eta ~,
\label{burgers}
\end{equation}
or, equivalently, the Kardar-Parisi-Zhang (KPZ) equation
\cite{2}
for the height $h$,
\begin{equation}
\frac{\partial h}{\partial t} = \nu\nabla^2 h +
\frac{\lambda}{2}\nabla_n h\nabla_n h
+ \eta ~,
\label{kpz}
\end{equation}
provide maybe the simplest continuum description of an
open driven nonlinear system exhibiting strong coupling features
such as dynamical scaling and pattern formation.
In (\ref{burgers}) and (\ref{kpz}) $\nu$ is a damping constant or viscosity
characterizing the linear diffusive term, $\lambda$ a coupling strength
for the nonlinear mode coupling or growth term, and $\eta$ a Gaussian
white noise driving the system into a stationary state and correlated
according to
\begin{equation}
\langle{\eta(x_n,t)\eta(x_n',t')}\rangle =
\Delta\prod_n\delta(x_n-x_n')\delta(t-t') ~,
\label{noise}
\end{equation}
characterized by the noise strength $\Delta$.

Notwithstanding the simple form of (\ref{burgers}) and (\ref{kpz}),
the driven Burgers equation introduced originally in order to model
aspects of turbulence and the KPZ equation providing the simplest
description of a growing interface, the morphology and scaling properties
embodied in (\ref{burgers}) and (\ref{kpz}) have been difficult to extract
and a full understanding of (\ref{burgers}) and (\ref{kpz}) remains
one of the important issues in nonequilibrium statistical physics
\cite{3}.

Besides perturbation theory in $\lambda$ \cite{4} which as regards the scaling 
properties provides the roughness and dynamic exponents
$(\zeta,z)=(1/2,3/2)$ in $d=1$, but, otherwise, is limited to an $\epsilon$
expansion about the (lower) critical dimension $d=2$, yielding a kinetic phase
transition above $d=2$ separating a weak coupling phase (the $\lambda = 0$
universality class) with exponents $(\zeta,z)=((2-d)/2,2)$ from a
strong coupling phase and to all orders in $\epsilon$ the exponents 
$(\zeta,z)=(0,2)$ on the phase line \cite{5}, nonperturbative 
methods include 1) in the $d=1$ case mapping to spin models \cite{6} and 
information
gained from lattice models \cite{7}, 2) mapping to directed polymers 
in combination
with replica methods \cite{3}, 3) mode coupling expansions
\cite{8}, and, 
most recently,
4) operator expansions yielding the strong coupling exponents
$(\zeta,z)=(2/5,8/5)$ in $d=2$ and $(\zeta,z)=(2/7,12/5)$ in $d=3$,
corresponding to skewness in the height distribution \cite{9}.

In a recent series of papers \cite{10}
we advanced a nonperturbative
approach to the noisy Burgers equation which purports to 
elucidate both the morphology
and scaling properties of a growing interface in $d=1$. 
Arguing that the noise strength $\Delta$ is the relevant 
nonperturbative parameter, driving the system into a stationary
state, and thus circumventing 1) the limitations
of perturbation theory which is based on an effective expansion
in $\lambda^2\Delta/\nu^3$ assuming regularity in $\Delta$ 
\cite{4} and 2)
the self consistency assumptions underlying the mode coupling
approach \cite{8}, the method was
based on a weak noise saddle point approximation to the Martin-Siggia-Rose
functional formulation \cite{11} of the noisy Burgers equation (\ref{burgers}).
Importantly, the method yields coupled deterministic field equations
for the slope $u$ and a noise field $\varphi$ (characterizing $\eta$),
replacing the stochastic Burgers equation, admitting soliton solutions 
and as a result a many body formulation of the pattern 
formation of a growing interface 
in terms of a dilute gas of propagating solitons with superposed
linear diffusive modes. 
The canonical form of the approach also yields 1) the gapless soliton
dispersion, $E\propto\lambda p^z$, $z=3/2$, and gapful diffusive mode
dispersion $\omega=-i\nu(k^2+(\lambda u/2\nu)^2)$, $u$ is the soliton
amplitude, 2) recovers the scaling exponents $(\zeta,z)=(1/2,3/2)$ and
an expression for the scaling function, and 3) associates the Burgers
universality  class with the leading gapless soliton excitation.

In the present letter we develop a general canonical phase space approach
to a stochastic Langevin equation of the Burgers type with additive
white noise. This method which emerged from our previous studies alluded to
above allows us to discuss and in some cases derive
the stationary and time-dependent weak noise solutions of the
associated Fokker-Planck equation for the probability distributions.
In particular for the Burgers equation (\ref{burgers}) the time-dependent
and stationary distributions are given by
\begin{eqnarray}
&&P(u_n,T)\propto
\exp{\left[-\frac{1}{\Delta}S(u_n,T)\right]} ~,
\label{prob}
\\
&&P_{\text{st}}(u_n)\propto
\lim_{T\rightarrow\infty}\exp{\left[-\frac{1}{\Delta}S(u_n,T)\right]} ~,
\label{statprob}
\end{eqnarray}
where the action has the canonical (symplectic) form
\begin{equation}
S = \int_0^Td^dxdt
\left(p_n\frac{\partial u_n}{\partial t}-{\cal H}\right) ~,
\label{act}
\end{equation}
with Hamiltonian density 
\begin{equation}
{\cal H} = p_n\left(\nu\nabla^2 u_n +
\lambda u_m\nabla_m u_n -
(1/2)\nabla_n\nabla_m p_m\right) ~,
\label{ham}
\end{equation}
yielding the coupled Hamiltonian equations of motion
\begin{eqnarray}
\left(\frac{\partial}{\partial t} -
\lambda u_m\nabla_m\right)u_n
&&= \nu\nabla^2 u_n -
\nabla_n\nabla_m p_m ~,
\label{can1}
\\
\left(\frac{\partial}{\partial t} -
\lambda u_m\nabla_m\right)p_n
&&= - \nu\nabla^2 p_n\nonumber
\\
&&+
\lambda(p_n\nabla_mu_m-
p_m\nabla_nu_m) ~.
\label{can2}
\end{eqnarray}
The above equations allow an analysis of the time-dependent
distribution $P(u_n,T)$. In principle we have to solve the
canonical field equations (\ref{can1}) and (\ref{can2}), determining
the orbits in $p_nu_n$ phase space, and compute the action (\ref{act})
and thus $P(u_n,T)$ according to (\ref{prob}).

The general framework is developed along the following lines:
The Fokker-Planck equation pertaining to a general Langevin equation
with additive noise for the stochastic variable $q_i$ 
($i$ is a discrete and/or continuous index),
\begin{eqnarray}
&&\frac{dq_n}{dt} 
= -\frac{1}{2}F_n(q_l) + \eta_n ~,
\label{lan}
\\
&&\langle\eta_n(t)\eta_m(t')\rangle
=\Delta K_{nm}\delta(t-t') ~,
\label{noise2}
\end{eqnarray}
has the form, denoting $\nabla_n = \partial/\partial q_n$,
\begin{equation}
\frac{\partial P}{\partial t} = \frac{1}{2}
\nabla_n[\Delta K_{nm}\nabla_m P + F_n P] ~.
\label{fp}
\end{equation}
Searching for a solution of the form 
\begin{equation}
P\propto\exp{\left[-\frac{1}{\Delta}S\right]} ~,
\label{ac2}
\end{equation}
it is an easy task to show that to leading order in the noise strength
$\Delta$, the action $S(q_n,t)$ satisfies the Hamilton-Jacobi
equation \cite{12}
\begin{eqnarray}
\frac{\partial S}{\partial t} + H(q_n,\nabla_nS) = 0 ~,
\label{hj}
\end{eqnarray}
where the energy $E=H$ and the canonically conjugate momentum
$p_n=\nabla_nS$. The Hamiltonian $H$ has the general form
\begin{eqnarray} 
H = (1/2)(K_{nm}p_np_m-F_np_n) ~,
\label{ham2}
\end{eqnarray} 
implying the Hamiltonian equations of motion
\begin{eqnarray} 
\frac{dq_n}{dt} 
&&=K_{nm}p_m-\frac{1}{2}F_n ~,
\label{cangen1}
\\
\frac{dp_n}{dt} 
&&=\frac{1}{2}p_m\nabla_nF_m ~.
\label{cangen2}
\end{eqnarray}
Assuming for simplicity that $F_n\rightarrow 0$ for
$q_n\rightarrow 0$ the energy surfaces have the characteristic submanifold
structure depicted in Fig. 1.

The origo in phase space constitutes a hyperbolic stationary point defined
by the unstable zero-energy submanifold $p_n=0$, the {\em transient manifold},
and, assuming the existence of a stationary state, 
a stable submanifold defined by
$K_{nm}p_m - F_n$ orthogonal to $p_n$, the {\em stationary manifold}.
The stationary state is determined by orbits on the zero-energy manifolds
whose structure thus characterizes the nature of the stochastic problem. The
action $S$ and hence the distribution $P$ are given by
\begin{eqnarray}
S(q_n,T,q_n') &&= \int_0^T dt\left[p_n\frac{dq_n}{dt}-H\right] ~,
\label{act2}
\\
P(q_n,T,q_n') &&\propto \exp{[-S(q_n,T,q_n')/\Delta]} ~,
\label{dist}
\end{eqnarray}
where the orbit from $q_n'$ to $q_n$ is traversed in time $T$.
The stationary distribution $P_{\text{st}}(q_n)$ is thus obtained
in the limit $T\rightarrow\infty$ and $E=0$, assuming that
$E(T)\propto\exp{[-\text{const.}T]}$ for $ T\rightarrow\infty$,
\begin{eqnarray}
P_{\text{st}}(q_n)=\exp{\left[-\frac{1}{\Delta}
\int_0^\infty dtp_n\frac{dq_n}{dt}\right]}  ~,
\label{statdist}
\end{eqnarray} 
i.e., $P_{\text{st}}$ is determined by the {\em infinite time orbits on the
zero-energy manifold}.
\begin{figure}
\begin{picture}(100,150)
\put(0.0,-190.0)
{
\centerline
{
\epsfxsize=13cm
\epsfbox{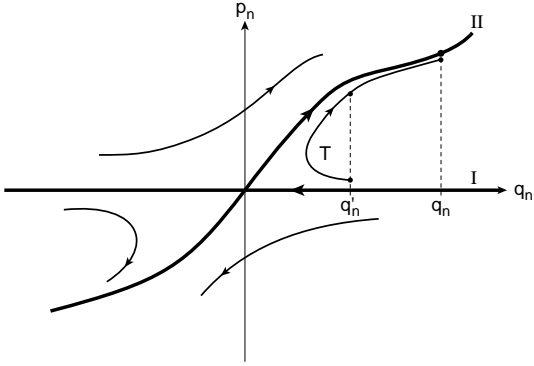}
}
}
\end{picture}
\caption{
Canonical phase space in the general case. The solid curves
indicate the zero-energy {\em transient} submanifold (I) and
{\em stationary} submanifold (II). The stationary saddle point is at the origin.
The finite time ($T$)
orbit from $q'_n$ to $q_n$ migrates to the zero-energy submanifold
for $T\rightarrow\infty$.
}
\end{figure}
This structure of phase space permits a simple non-stochastic,
deterministic discussion of the approach to the stationary
state of a damped noise-driven system in terms of dynamical system theory.
Referring to Fig. 1, consider an orbit from $q_n'$ to $q_n$ on the 
energy surface $E(T)$ traversed in time $T$. In order to attain the
stationary state $E(T)\rightarrow 0$ in the limit $T\rightarrow\infty$.
For $E\sim 0$ the initial part of the orbit moves close to the $p_n=0$
submanifold and from (\ref{cangen1}) is determined by
$dq_n/dt = -(1/2)F_n$
i.e., the deterministic noiseless version of the Langevin equation
(\ref{lan}). In the absence of noise the motion is transient and damped.
The orbit slows down near the stationary point (the origo in phase space)
before it picks up again and moves close to the other submanifold
$K_{nm}p_m-F_n\perp p_n$. This final part of the orbit terminating in $q_n$
at time $T$ thus corresponds to the establishment of the stationary 
state. Markovian behavior, i.e., the independence of the initial configuration,
is associated
with the long (infinite) {\em waiting time} near (at) the stationary point.

In the special case $K_{nm}=\delta_{nm}$ and $F_n=\nabla_n\Phi$,
corresponding to an effective fluctuation-dissipation theorem and
an underlying free energy $\Phi$, the energy (\ref{ham2}) and equations of
motion (\ref{cangen1}) and (\ref{cangen2}) are consistent with the zero-energy
submanifolds $p_n=0$ and $F_n=p_n$ and yield the equilibrium
distribution $P_{\text{st}}(q_n)\propto\exp{[-\Phi/\Delta]}$.
In the general case a determination of the time-dependent and stationary
distributions require a knowledge of the energy submanifolds in 
combination with a solution of the canonical equations in order to 
determine the orbits.

The noisy Burgers and KPZ equations (\ref{burgers}) and (\ref{kpz})
fall
within the scope of the general framework summarized above.
With the identification $q_i(t)\rightarrow u_n(x_m,t)$, 
$K_{nm}\rightarrow\nabla^2\prod_m\delta(x_m-x_m')$, and
$F_n\rightarrow -2(\nu\nabla^2u_n+\lambda u_m\nabla_mu_n)$,
we obtain  (\ref{prob})-(\ref{can2}). Note that the canonical
momentum, the `noise field' $p_n$, is essentially a `slaved variable'.

In the case $d=1$ which is the basis for our discussion here, the equations
(\ref{can1}) and (\ref{can2}) reduce
to the form
\begin{eqnarray}
\left(\frac{\partial}{\partial t}-\lambda u\nabla\right)u
&&=\nu\nabla^2 u - \nabla^2p ~,
\label{can1d1}
\\
\left(\frac{\partial}{\partial t}-\lambda u\nabla\right)p 
&&=-\nu\nabla^2 p ~.
\label{can1d2}
\end{eqnarray}
The distribution  (\ref{prob}) is determined by the action (\ref{act})
which has the general structure
\begin{eqnarray}
S = S_{\text{st}}(u) + S_{\text{skew}}(u,T) + S_{sol}(u,T) ~.
\label{genact}
\end{eqnarray}

The {\em stationary distribution} $P_{\text{st}}(u)$ given by 
$S_{\text{st}}(u)$ is 
easily found by noting that (\ref{can1d1}) and (\ref{can1d2})
coincide on the zero-energy submanifold $p=2\nu u$
(the energy density ${\cal H}$ becomes a total derivative yielding $E=0$).
Inserting $p=2\nu u$  and $E=0$ in (\ref{prob}) and (\ref{act}) in the 
limit $T\rightarrow\infty$ and integrating over time we obtain the
the  symmetric Gaussian stationary distribution \cite{13}
\begin{eqnarray}
P_{\text{st}}(u)\propto\exp{\left[-(\nu/\Delta)\int dx~u(x)^2\right]} ~.
\label{stadis2}
\end{eqnarray}
Also, setting $p=2\nu(u+\delta u)$ we find to leading order in 
$\delta u$ $(\partial/\partial t-\lambda u\nabla)\delta u=
\nu\nabla^2\delta u$.
Noting that $\partial/\partial t - \lambda u\nabla$ is invariant
under the  Galilean transformation:
$x\rightarrow x-\lambda u_0t$, $u\rightarrow u +u_0$, and  choosing 
an instantaneous frame with vanishing $u$,
$\delta u\propto\exp{[-\nu k^2 t]}$,
implying that the orbits approach the zero-energy stationary 
submanifold $p=2\nu u$.

The {\em long time skew distribution} $P_{\text{skew}}(u,T)$ is determined by
$S_{\text{skew}}(u,T)$. For $\lambda=0$ we obtain in wave number
space the symmetric contribution, $u_k=\int dx \exp(-ikx)u(x)$,
\begin{eqnarray}
S_{\text{skew}}^0(u_k,T) = -\nu\int\frac{dk}{2\pi}|u_k|^2\exp[-2\nu k^2T] ~,
\end{eqnarray}
defining a crossover time  $T_{\text{co}}\propto 1/\nu k^2$.
For a finite system $k\propto 1/L$ ($L$ is the system size), i.e.,
$T_{\text{co}}\propto L^2/\nu$, yielding the dynamic exponent $z=2$ 
in accordance with the diffusive mode contribution.

For $\lambda\neq 0$ and for large $T$ approximating the orbit
{\em close to the manifold} by an orbit {\em on the manifold},
inserting $p=2\nu u$ in (\ref{can1d1}), we obtain the deterministic
Burgers equation with $\eta = 0$ and viscosity $-\nu$ which can be
solved by means of the Cole-Hopf transformation \cite{2}.
Thus setting $u=\nabla h$ and $h=-(2\nu/\lambda)\ln{w}$ yields
the diffusion equation $\partial w/\partial t=-\nu\nabla^2w$
for $w$ which is solved by means of the Green's function
$G_{x}(T)=[4\pi\nu T]^{-1/2}\exp{[-x^2/4\nu T]}$. We obtain
$S_{\text{skew}}(u,T) = \nu\int dx~u'(x)^2$,
where $u=\nabla h$ and $u'=\nabla h'$ are related according to
the nonlinear expression
\begin{eqnarray} 
\exp[-(\lambda/2\nu)h']
=
\int dx' G_{x-x'}(T)
\exp[-(\lambda/2\nu)h] ~,
\label{ch}
\end{eqnarray}   
giving rise to the distribution
\begin{eqnarray}
P(u,T)\propto 
P_{\text{st}}(u)\exp\left[\frac{\nu}{\Delta}\int dx~u'(x)^2\right] ~,
\end{eqnarray}
which by inspection is skew.
To order $\lambda$ in wave number space we also have, setting
$G_{k,T}=\exp[-\nu k^2 T$],
\begin{eqnarray}
P(u_k,T)\propto 
P_{\text{st}}(u_k)P_{\text{skew}}^0(u_k,T)P_{\text{skew}}^\lambda(u_k,T) ~,
\end{eqnarray}
where $P_{\text{skew}}^0\propto\exp[-S_{\text{skew}}^0/\Delta]$,
$P_{\text{skew}}^\lambda\propto\exp[-S_{\text{skew}}^\lambda/\Delta]$,
and 
\begin{eqnarray}
&&S_{\text{skew}}^\lambda(u_k,T)= 2\lambda\int\frac{dk}{2\pi}\frac{dk'}{2\pi}
F_{k,k',T}
\nonumber
\\
&&F_{k,k',T} = [G_{k,2T}-G_{k,T}G_{k+k',T}G_{k',T}]
u_ku_{-k-k'}h_{k'} ~,
\nonumber
\\
\label{skew4}
\end{eqnarray}
exhibiting skewness.

The {\em short time distribution} $P_{\text{sol}}(u,T)$ given by 
$S_{\text{sol}}(u,T)$,
corresponding to an orbit off the zero-energy manifold, is determined 
by the soliton-diffusive mode contribution discussed in \cite{10}.
For a single soliton with boundary values $u_+$ and $u_-$ the propagation
velocity $v$ is given by the soliton condition
\begin{equation}
u_++u_-=-2v/\lambda ~.
\label{solcon}
\end{equation}
However, only the {\em left hand} soliton ($u_+<u_-$) carries nonvanishing
energy, momentum and action according to the assignment
$E=(2/3)\nu\lambda(u_+^3-u_-^3)$, $\Pi=\nu(u_+^2-u_-^2)$, and
$S=(1/6)\nu\lambda T|u_+-u_-|^3$ (note the Galilean invariance of $S$).
The action of a multi-soliton configuration constituting a growing 
interface is then given by
\begin{equation}
S_{\text{sol}}(u,T) = \frac{1}{6}\lambda\nu T\sum_{\text{lhs}}|u_+-u_-|^3 ~,
\label{acsol}
\end{equation}
where summation is over {\em left hand} solitons (lhs) only.

Owing to the constraint imposed by the soliton condition (\ref{solcon})
and the non-integrability of the equation of motion we can only give a
qualitative discussion of $P_{\text{sol}}$. Since the saturation width of an
interface is a finite size effect time scale separation only occurs for
a finite system. Noting that the propagation of solitons and the imposition
of periodic (or bouncing) boundary conditions, in order to ensure growth
in $h$, endows the velocity with a scale, i. e., $v\sim L/T$, we obtain,
inserting $u_++u_-=-2v/\lambda\sim u_+-u_-$ in (\ref{acsol}) and from
$P_{\text{sol}}\sim\exp[-S_{\text{sol}}/\Delta]$ the soliton crossover time
\begin{equation}
T_{\text{co}}^{\text{sol}}\propto(1/\lambda)(\nu/\Delta)^{1/2}L^{3/2} ~,
\label{scot}
\end{equation}
which is also is consistent with the dimensionless argument
$\lambda(\Delta/\nu)^{1/2}t/x^{3/2}$ in the scaling function for
the slope correlations discussed in \cite{8,10}.

In the short time regime $T\ll T_{\text{co}}^{\text{sol}}$ the soliton configurations
contribute to $P$. Noting that 
$|u_+-u_-|^3\sim(uL)^{3/2}(T\lambda)^{-3/2}\sim
h^{3/2}(T\lambda)^{-3/2}$ we obtain, inserting in (\ref{acsol})
($h$ is measured relative to the mean height)
\begin{equation}
P_{\text{sol}}(h,T)\propto
\exp[-(\nu/\Delta)(1/\lambda T)^{1/2}h^{3/2}] ~,
\end{equation}
in accordance with the directed polymer-replica based result \cite{3}
and the exact results for the asymmetric exclusion model \cite{14}.
The skewness of the distribution then arises from the bias in the 
statistical weight $\exp[-S/\Delta]$ assigned to the {\em left}
and {\em right hand} solitons giving rise to a predominance of
{\em right hand} solitons ($S=0$), corresponding to relative forward 
growth. In the long time regime, $T\gg T_{\text{co}}^{\text{sol}}$, 
the soliton contribution vanishes and only the diffusive modes and
their interactions contribute to $P$.

In this letter we have outlined a novel  canonical phase space
approach to Langevin equations with additive noise; details will appear
elsewhere. In addition to
providing insight from dynamical system theory the method also yields a
calculational tool for the determination of the weak noise 
probability distributions. In particular we have applied the 
method to the noisy Burgers equation in $d=1$ and derived expressions
for the skew finite time probability distribution. In the short
time regime our heuristic result is in agreement with the directed
polymer-replica method. 

Discussions with M. Kosterlitz, T. Hwa, J. Hertz, P. Cvitanovi\'{c},
K.B. Lauritsen, and A. Svane
are gratefully acknowledged.

\end{multicols}

\begin{references}
\bibitem{1}
D. Forster, D.~R. Nelson, and M.~J. Stephen, Phys. Rev. Lett. {\bf 36},  867
(1976). 
\bibitem{2}
M. Kardar, G. Parisi, and Y.~C. Zhang, Phys. Rev. Lett. {\bf 56},  889  (1986).
\bibitem{3}
T. Halpin-Healy and Y.~C. Zhang, Phys. Rep. {\bf 254},  215  (1995).
\bibitem{4}
E. Frey and U.~C. T\"{a}uber, Phys. Rev. E {\bf 50};
Rev. E {\bf 51},  6319  (1995).
\bibitem{5}
M. L\"{a}ssig, Nucl. Phys. B {\bf 448}, 559 (1995).
\bibitem{6}
L.~-H. Gwa and H. Spohn, Phys. Rev. Lett. {\bf 68}, 725 (1992);
H.~C. Fogedby, A.~B. Eriksson and L.~V. Mikheev,
Phys. Rev. Lett. {\bf 75}, 1883 (1995).
\bibitem{7}
H.~K. Janssen and B. Schmittman,
Z. Phys. B {\bf 63}, 517 (1986);
B. Derrida, M.~R. Evans, V. Hakim, and V. Pasquier, J. Phys. A {\bf 26},
1493 (1993)
\bibitem{8}
E. Frey, U.~C. T\"{a}uber, and T. Hwa,
Phys. Rev. E {\bf 53}, 4424 (1996).
\bibitem{9}
M. L\"{a}ssig, Phys. Rev. Lett. {\bf 80}, 2366 (1998);
Chen-Shan Chin and M. den Nijs, cond-mat/9810083.
\bibitem{10}
H.~C. Fogedby, Phys. Rev. E {\bf 57},  2331  (1998);
Phys. Rev. E {\bf 57},  4943  (1998);
Phys. Rev. Lett. {\bf 80},  1126  (1998);
cond-mat/9809238.
\bibitem{11}
P.~C. Martin, E.~D. Siggia, and H.~A. Rose, Phys. Rev. A {\bf 8},  423  (1973);
R. Baussch, H.~K. Janssen, and H. Wagner, Z. Phys. B {\bf 24},  113  (1976).
\bibitem{12}
L. Landau and E. Lifshitz, {\em Mechanics} (Pergamon Press, Oxford, 1959);
\bibitem{13}
D.~A. Huse, C.~L. Henley, and D.~S. Fisher,
Phys. Rev. Lett. {\bf 55}, 2924 (1985).
\bibitem{14}
B. Derrida, J.~L. Lebowitz,
Phys. Rev. Lett. {\bf 80}, 209 (1998).
\end{references}
\end{document}